   \newcommand{\be}[0]{\begin{equation}}
   \newcommand{\ee}[0]{\end{equation}}
   \newcommand{\ba}[0]{\begin{eqnarray}}
   \newcommand{\ea}[0]{\end{eqnarray}}

\newcommand{\scriptF}{\mathcal{F}}
\newcommand{\scriptG}{\mathcal{G}}
\newcommand{\MSb}{{\rm \overline{MS}}}
\newcommand{\MSbPS} {{\rm \overline{MS}PS}}
\newcommand{\LtMSb}{\tilde{\Lambda}_{\MSb}}
\newcommand{\LMSb}{\Lambda_{\MSb}}
\newcommand{\alphazero}{\overline{\alpha}_0}
\newcommand{\refeq}[1]{Eq. (\ref{eq:#1})}
\newcommand{\R}{{\mathcal R}}
\newcommand{\RS}{{\mathrm {RS}}}

\documentclass[12pt]{article}
\usepackage{epsfig} \usepackage{amssymb} \usepackage{amsfonts}

\begin{document}

\Large
\hfill\vbox{\hbox{IPPP/04/77}
            \hbox{DCPT/04/154}}

\nopagebreak

\vspace{0.75cm}
\begin{center}
\LARGE
{Resummation of Large Logarithms within the Method of Effective Charges}
\vspace{0.6cm}
\Large

M.~J.~Dinsdale\footnote{email:{\tt m.j.dinsdale@durham.ac.uk}} and C.~J.~Maxwell\footnote{email:{\tt c.j.maxwell@durham.ac.uk}}

\vspace{0.4cm}
\large
\begin{em}
Institute for Particle Physics Phenomenology, University of Durham,
South Road, Durham, DH1 3LE, UK
\end{em}

\vspace{1.7cm}

\end{center}
\normalsize
\vspace{0.45cm}

\centerline{\bf Abstract}
\vspace{0.3cm}
We show how the resummation of large logarithms can be incorporated into the method
of effective charges.  As an example, we apply this approach to the event shape variables
thrust and heavy-jet mass in $e^{+}e^{-}$ annihilation.  We find that, although the resummation
creates problems with the behaviour of the effective charge in the 2-jet limit,
smaller power corrections are required to fit the data compared to the standard approach.  In addition, increasing
the logarithmic accuracy reduces the size of the power corrections further.
We also consider ``predicting''
the sub-leading logarithms in the $\overline{MS}$ scheme, obtaining surprisingly good results for the first few NLL terms.

\newpage
\section{Introduction}

Over the past decade, QCD has moved on from giving a qualitatively good description of strong interaction physics
and entered an era of real quantitative tests.  A wealth of precision data has appeared, particularly from the experiments at LEP
and HERA, and it is a challenge for theorists to attain a similar precision in their calculations.
A major stumbling block in this attempt is the computational complexity of QCD perturbation theory.
In practice this means that we are limited to using the first 2 to 3 terms of the perturbation series in the strong coupling $\alpha_S$.
The relatively large value of $\alpha_S$ 
compared, say, to $\alpha_{QED}$, means that this unavoidable truncation of the perturbation series seriously limits the accuracy of perturbative
QCD (pQCD) calculations.  This problem is further complicated by the spurious dependence of fixed-order results on the renormalization scheme (RS)
and scale ($\mu$), which together can be termed the renormalization prescription (RP) \cite{Fischer:1997bs}.  This dependence, which does not afflict
the exact results, requires us to choose some suitable RP to make pQCD predictions.  There are in principle infinitely many ways to do this; by far the most
common in practice is to choose the $\MSb$ RS and set $\mu$ equal to some physical scale of the process.  For brevity, let us
refer to this approach as $\MSbPS$ (for $\MSb$ scheme with a physical scale).  Despite its popularity, this choice of RP is not
supported by any theoretical argument; the motivation for it is purely empirical, namely the evident success of standard pQCD phenomenology,
where this approach is almost invariably taken.  However, there also exist several theoretical proposals for choosing so-called ``optimized'' RPs,
tuned to the observable at hand using information that is already present in the perturbative calculations.  This is in contrast
to the idea of selecting a ``physical'' scale which requires an additional judgement to be made for each observable outside the well-defined framework of pQCD.  Of course,
the choice to use a particular ``optimized'' approach is also an additional judgement, but the point is that these approaches are supposed to pick
the RP for {\it each observable} automatically once the optimization method itself is specified.

One such optimization method, the method of effective charges (ECH) \cite{Grunberg:1982fw}, was recently applied in Ref.\cite{r12} to the description
of event shape means.  The authors found that with this choice of RP, next-to-leading order (NLO) pQCD could describe the data very well
even {\it without} the introduction of power-suppressed non-perturbative effects.  This surprising result suggests that it might be interesting to
apply ECH also to the distributions of these event shapes.  Such analyses have been performed \cite{r9,Abreu:2000ck},
but unfortunately in this case ECH suffers a restricted range of
validity due to kinematical end-points and logarithmic enhancements.  This latter problem affects also the $\MSbPS$ treatment of
event shape distributions, where it can be handled by a resummation
of large logs.  A way to implement this within ECH was presented in Ref.\cite{r15}, but a systematic recipe for carrying out the ECH analysis of
a general observable known to, say, next-to-leading logarithmic accuracy, is not available in the literature.  In this paper
we present such a recipe.  We then go on to test it by applying it to the problem of describing event shape distributions in the 2-jet limit.

The plan of this paper is as follows:  we  briefly review ECH in Section \ref{se:ECH}.
In Section \ref{se:resummation} we outline a general procedure for obtaining an all-orders resummed ECH result starting from an
all-orders resummed $\MSbPS$ perturbation series.
Section \ref{se:ESV} gives our application of this to the distributions of the event shape variables thrust and heavy-jet mass
in $e^+e^-$ annihilation.  Section \ref{se:conclusions} contains our conclusions.

\section{The Method of Effective Charges}
\label{se:ECH}

This section briefly summarises the method of effective charges; for more detailed developments see e.g Refs.\cite{Grunberg:1982fw,r10}.
Consider a quantity normalized so that its perturbation series takes the form
\be
\label{eq:R}
a(\mu,\RS)+r_1(\mu/Q,\RS) a^2(\mu,\RS) + r_2(\mu/Q,\RS) a^3(\mu,\RS) + \cdots,
\ee
with $a=\alpha_S/\pi$.  Such a quantity could either be the coupling defined in some RS with $Q$ as the renormalization scale
or a suitably normalized physical observable $\R(Q)$ depending on a single energy scale $Q$.  Indeed, these possibilities are not mutually exclusive.
Given some observable of the form \refeq{R}, one can define an RP such that all the $r_n$ vanish; in this RP, $\R=a$,
so the observable is equal to the coupling.  Such an observable/coupling is called an {\it effective charge} \cite{Grunberg:1982fw}.  A typical
example is the R-ratio in $e^+e^-$ annihilation, with which one can associate an effective charge $\R(Q)$ via
\be
R_{e^+e^-}=N_c \left( \sum_f{Q^2_f} \right) (1 + \R(Q)).
\ee

Whatever the quantity represented in \refeq{R}, it will be independent of the RP in which the expansion is performed.
Moreover, when the expansion is truncated at some order, the variation in this partial sum due to a change of RP is always
one order higher in $a$.  This implies specific relations between the $r_n$ and the RP \cite{Stevenson:1981vj}.

The dependence of the coupling $a$ defined in some RS on $\mu$ is described by the beta-function
\be
\label{eq:beta-function}
\frac{da(\mu,\RS)}{d\ln(\mu)}=\beta(a)=-b a^2 (1 + c a + c_2 a^2 + c_3 a^3 + \cdots),
\ee
where $b$ and $c$ are independent of the RS, and the $c_i, \; i\geq 2$ can be taken to label the RS, along
with the scale parameter $\Lambda$ which we introduce below \cite{Stevenson:1981vj}.
Restricting $a$ to the subset of couplings which are also effective charges, the same equation is conventionally written as
\be
\label{eq:EC-beta-function}
\frac{d\R(Q)}{d\ln(Q)}=\rho(\R)=-b \R^2 (1 + c \R + \rho_2 \R^2 + \rho_3 \R^3 + \cdots),
\ee
where $b$ and $c$ are now independent of the choice of effective charge, whereas the $\rho_n$ depend on this choice.

Because $\rho$ describes a functional relation between two physical quantities, namely $\R$ and its
energy derivative, it is automatically independent of any choice of RP we make in calculating it.
Clearly, the $\rho_n$ must also be similarly RP-independent.  This means that truncating $\rho$ at some
fixed order in perturbation theory, and calculating an approximation for $\R$ by integrating the
corresponding approximate effective charge beta-function will give RP-independent results.
Another way of looking at this is to say that the method of effective charges involves a specific choice of RP
(i.e. the RP where $\R=a$), so that the energy evolution of the observable is identical to the beta-function evolution
of the coupling.

Now, to obtain a perturbative approximation to $\rho$, consider setting $\mu=Q$ in \refeq{R}
(which is allowable as the full sum is independent of $\mu$), and then differentiating
with respect to $\ln Q$.  This gives
\be
\label{eq:rho-from-diffs}
\rho(\R) \equiv \frac{d\R}{d\ln Q} = \left. \frac{\partial \R}{\partial a} \frac{d a(Q)}{d\ln Q} \right |_{a=a(\R)} = \left. \frac{\partial \R}{\partial a} \beta(a) \right |_{a=a(\R)}.
\ee
Expanding this order-by-order in $\R$ and comparing to \refeq{EC-beta-function} gives expressions for
the $\rho_n$ as
multinomials in the $r_n$ and the $c_n$; for example, the first two are
\begin{eqnarray}
\label{eq:rho_n}
\rho_2 & = & c_2 + r_2 - r_1 c - r_1^2 \\ \nonumber
\rho_3 & = & c_3 + 2r_3 - 4r_1 r_2 - 2 r_1 \rho_2 - r_1^2 c + 2r_1^3.
\end{eqnarray}

Suppose we have performed a NLO calculation of $\R$ (in some arbitrary RP).  This provides us
with $r_1$, but not $r_2$ or any higher-order coefficients.  So as this stage, our best approximation
to $\rho(\R)$ is simply the universal form $\rho(\R)=-b\R^2(1+c\R)$.  This can therefore be
termed the ``NLO'' approximation to $\rho$.  Given the $r_i$ and $c_i$ for $i=1..n$, we can calculate
the $\rho_i$ to the same order, obtaining a ${\mathrm N^n}$LO approximation.

Having constructed an approximation for $\rho$ it just remains to relate this to the effective charge
itself and hence to the corresponding observable.  This can be accomplished by integrating
\refeq{EC-beta-function} using asymptotic freedom as a boundary condition.  Note however that \refeq{EC-beta-function}
only fixes $\R$ up to an arbitrary rescaling of $Q$.  This is necessary, as the NLO $\rho$ is the same for {\it all}
effective charges, and they certainly should not all have the same NLO predictions.  Similarly, \refeq{beta-function}
only fixes $a$ up to a rescaling of $\mu$.  

The missing information
needed to fix the scale (corresponding to the single free parameter of massless QCD) can be conveniently
provided by a parameter $\Lambda$ associated to each RS and each effective charge which will
arise as a constant of integration.  It might seem contradictory to describe
$\Lambda$ as simultaneously differentiating between effective charges and acting as {\it the} free parameter of QCD.
The resolution of this apparent paradox is the so-called Celmaster-Gonsalves relation \cite{Celmaster:1979km},
which relates $\Lambda$ parameters
for different effective charges (or indeed different RS's) {\it exactly} using only their NLO coefficients computed in some
reference RP
\be
\label{eq:C-G}
\Lambda_{\R} e^{-r_1/b} = \Lambda_{\R '} e^{-r_1 '/b}.
\ee
(It does not matter what RP is used for the calculation of these $r_1$ coefficients as a change of RP just adds
a constant to $r_1$ which clearly drops out of this equation.)  We can therefore take any of these $\Lambda$'s
as the fundamental (dimensional transmutation) parameter of QCD and obtain any other $\Lambda$ from it given
only NLO information.  Conventionally we choose $\LMSb$.  However, there is a subtlety here related to the way that
$\LMSb$ is traditionally defined, so that to avoid confusion we refer to the $\LMSb$ we 
are discussing here as $\LtMSb$ \cite{Stevenson:1981vj}.  This is related to the more conventional $\LMSb$ \cite{Buras:1977qg} via
\be
\label{eq:tilde}
\LtMSb=\left( \frac{2c}{b} \right)^{-c/b} \LMSb.
\ee
After integrating \refeq{EC-beta-function} we arrive at
\be
\label{eq:blogQ}
b \ln \frac{Q}{\Lambda_\R} = \int^{\infty}_{\R(Q)} \frac{dx}{x^2 (1+cx)} + \int_0^{\R(Q)} dx \left [ \frac{-1}{x^2(1+cx+\rho_2 x^2 + \cdots)} + \frac{1}{x^2(1+cx)} \right].
\ee
The first integral is independent of the particular effective charge under consideration, whereas the second is not.  Exponentiating gives
\be
\label{eq:LambdaRfromFG}
\frac{\Lambda_{\R}}{Q} = \scriptF(\R(Q)) \scriptG(\R(Q))
\ee
where $\scriptF$ is universal and comes from the first integral in \refeq{blogQ}
\be\
\scriptF(\R) = e^{-1/b\R} \left(1 + \frac{1}{c\R} \right) ^ {c/b},
\ee
and $\scriptG$ comes from the second integral and depends on the effective charge,
\be
\label{eq:curlyG}
\scriptG(\R) = \exp \left[ -\int_0^{\R(Q)}{dx \frac{1}{\rho(x)} + \frac{1}{bx^2(1+cx)}} \right].
\ee
Approximating $\rho$ by its NLO form gives $\scriptG_{\rm NLO}=1$.
Converting from $\Lambda_{\R}$ to $\LMSb$ using \refeq{C-G} and \refeq{tilde} gives
\be
\label{eq:extracter}
\LMSb = Q \scriptF(\R(Q)) \scriptG(\R(Q)) e^{-r_1(\overline{\mathrm MS}, \mu=Q)/b} \left( \frac{2c}{b} \right)^{c/b}.
\ee
This equation allows us to extract values of $\LMSb$ directly from the observed values of $\R$, or to make predictions
for $\R$ (by solving the implicit equation e.g. iteratively).

One way to compare this with the more standard approach of truncating the series for $\R$ in some fixed RP
is to write an ``effective'' effective charge beta-function describing the energy evolution of $\R$ implied by
this standard approach.  Let us call this function $\bar{\rho}$.  It can be computed from \refeq{rho-from-diffs}
using the truncated relation between $\R$ and $a$.  It always agrees with the true $\rho$ up to the order (NLO, NNLO,...)
to which $\R$ has been calculated, but it also includes terms of all higher orders in $\R$ which depend on the RP
we chose to perform our truncation in.  If the RP is such that $r_1$ is large, these higher order terms are also large,
so the ECH predictions will differ radically from the standard prediction {\it in that RP}.  For many $e^+e^-$ jet observables
$r_1$ is indeed large in the $\MSbPS$ RP.

\section{Resummation of Logarithms in the ECH Beta-Function}
\label{se:resummation}

It is commonly stated that the method of effective charges is inapplicable to
exclusive quantities such as event shape distributions.  The idea is that the dependence of the physical quantity
on multiple scales invalidates the derivation of the ECH beta-function as presented here in Section \ref{se:ECH}.  However, as pointed out in \cite{r10}, this is not
really the case.  Given an observable $\mathcal{R}=\mathcal{R}(Q_1,Q_2,...,Q_n)$ depending on $n$ scales, one can simply re-express it as
$\mathcal{R}=\mathcal{R}(Q_1,Q_2/Q_1,...,Q_n/Q_1) \equiv \mathcal{R}_{x_2,...,x_n}(Q_1)$.  Here the $x_i \equiv Q_i/Q_1$
are {\it dimensionless} quantities that can be thought of as labelling the effective charge which is now a function of one single
dimensionful scale $Q_1$.  We can then write an effective charge beta-function for this $\mathcal{R}$ describing the energy evolution
of our observable for fixed values of the ratios $x_i$.  Still, this formal manipulation cannot tell us whether the $\rho$ function we
arrive at in this way will be well approximated by its NLO terms, which is what we require for most fixed-order phenomenological
applications, given the current state of the art in perturbative QCD calculations.  One reason in particular why this
might not be the case is if some of the $x_i$ become large - typically this
leads to powers of large logs $L_i=\log(x_i)$ enhancing the coefficients $r_n$ in the perturbation series for $\mathcal{R}$, and hence also the $\rho_n$
in the corresponding $\rho(\cal{R})$ function.  A common situation is that more logs appear as the order of perturbation theory is increased, so that
for a sufficiently large $L$ the terms all become of similar magnitude.
This invalidates both the NLO (universal) approximation for $\rho(\cal{R})$ and
NLO $\MSbPS$.  However, in the latter case, there is a well-known way out.  If the leading powers of the logs can be identified as having some
simple dynamical origin, we may be able to calculate them {\it to all-orders} and then effect a resummation, extending the validity of our
perturbative results into the large $L$ region.  This suggests that essentially the same trick might work for the $\rho$ function.
In this section we describe how to accomplish this, expanding on ideas in Ref.\cite{r15}; an example of the phenomenological application
of these ideas to event shapes is presented in the next section.

Our approach will be to start with some resummed result for the observable of interest calculated by conventional means.
As an example, consider an observable of the form
\be
O(L) = L A_{LL}(aL) +  A_{NLL}(aL) + a A_{NNLL}(aL) + \cdots\;,
\ee
where $a$ is as usual the couplant and $L$ is the large logarithm which this expression resums.
The subscripts ``LL'', ``NLL'' and ``NNLL'' stand for
leading log, next-to-leading log and next-to-next-to-leading log respectively.  We can
relate this observable to an effective charge
\be
O(L) = r_0(L) \mathcal{R}(L) = r_0(L) (a + r_1(L) a^2 + r_2(L) a^3 + \cdots).
\ee
Here $r_0$ is the leading order coefficient, whose large $L$ behaviour is $r_0 \sim L^2$.
The $r_n$ can now be expanded in powers of the large log $L$.  In this case their leading behaviour is $r_n \sim L^n$ and we can write
\begin{equation}
r_n = r_n^{\rm LL} L^n + r_n^{\rm NLL} L^{n-1} + \cdots\;,
\end{equation}
so the structure of the $\rho_n$ as illustrated in Eq.(\ref{eq:rho_n}) implies that
\begin{equation}
\rho_n = \rho_n^{\rm LL} L^n + \rho_n^{\rm NLL} L^{n-1} + \cdots\;.
\end{equation}
Because $L$ is a logarithm of a physical quantity, this expansion of the $\rho_n$ is RP-independent
We can thus define resummed, RP-independent approximations to $\rho(\cal{R})$ such as
\ba
\rho_{\rm LL}(\mathcal{R})& = &-b \mathcal{R}^2 (1 + c \mathcal{R} + \sum_{n=2}^{\infty} \rho_n^{\rm LL} L^n \mathcal{R}^n ) \\
\rho_{\rm NLL}(\mathcal{R})& = &-b \mathcal{R}^2 (1 + c \mathcal{R} + \sum_{n=2}^{\infty} (\rho_n^{\rm LL} L^n + \rho_n^{\rm NLL} L^{n-1}) \mathcal{R}^n )\;.
\ea
and so on.  These can be calculated order-by-order using the relations between the $\rho_n$ and $r_n$, the first few of which
are shown in Eq.(\ref{eq:rho_n}).  Alternatively, we can apply a numerical procedure to extract our desired $\rho$ function
from $\mathcal{R}$ calculated to similar logarithmic accuracy.  This is particularly simple for $\rho_{\rm LL}$, as can be seen
by considering the $\bar{\rho}$ corresponding to $\cal{R}_{\rm LL}$ with one-loop beta-function
\be
\label{eq:rhobarLL}
\bar{\rho}_{\rm LL}(x) = \beta(a) \frac{d\cal{R}_{\rm LL}}{da} = -ba^2 \frac{d\cal{R}_{\rm LL}}{da},
\ee
with $a$ chosen such that $\mathcal{R}_{\rm LL}(a)=x$.  The perturbative coefficients of this $\bar{\rho}$ function
can be obtained from the expressions for the $\rho_n$, of which the first two are shown in Eq.(\ref{eq:rho_n}),
using $c=0$,$\;c_i=0$ and $r_n = r_n^{\rm LL} L^n$.  But then it is easy to see
that the coefficients we obtain are proportional to $L^n$ and moreover identical to the coefficients of $\rho_{\rm LL}$ (because
adding the sub-leading terms in $\beta(a)$ and $\mathcal{R}$ only affects the $\bar{\rho}_n$ at next-to-leading logarithmic accuracy).  In other
words $\bar{\rho}_{\rm LL}$, defined as above, is {\it equal} to $\rho_{\rm LL}$ except for the $c \mathcal{R}$ term which can easily
be added (in some sense this term is NLL as it
is ${\rm O}(L^{-1})$ in the large $L$, fixed $\mathcal{R}L$ limit, but we include it in $\rho_{\rm LL}$ as it is obviously present in the full expression, and this avoids
having to modify \refeq{curlyG}).  Thus $\rho_{\rm LL}(\mathcal{R})$ can be calculated to arbitrary accuracy for a given $\mathcal{R}$ by numerically inverting
$\mathcal{R}_{\rm LL}(a)$ to obtain the corresponding $a$, then substituting this $a$ into Eq.(\ref{eq:rhobarLL}).

Going beyond leading log accuracy things become slightly more complicated, because the $\bar{\rho}$ functions pick up terms at lower logarithmic accuracy that
do not appear in our resummed approximations.  For example, at NLL we will have $r_1=r_1^{\rm LL} L + r_1^{\rm NLL}$, so that $\bar{\rho}_2$ will contain not only
$L^2$ and $L^1$ terms, but also $L^0$ terms.  These are not included in our definition of $\rho_{\rm NLL}$, and indeed they must not be, as they are affected
by the addition of the remaining missing terms in $\mathcal{R}$ and hence are RP-dependent.  However, it is still the case that the NLL terms in $\bar{\rho}_{\rm NLL}$
are unchanged by adding sub-leading terms (in $\beta(a)$ and $\mathcal{R}$), and are therefore identical to the corresponding terms in $\rho_{\rm NLL}$
(assuming that the $\bar{\rho}$ functions are defined with
beta-functions having sufficiently many terms to make this true, e.g. $\beta(a)=-ba^2(1+ca)$ for the NLL case).
So, truncating $\bar{\rho}$ by numerically taking limits ($L \to \infty$ with $L\cal{R}$ fixed) allows us to extract the LL and
NLL terms.  The generalization to higher logarithmic accuracy is straightforward.


Some physical quantities might have more divergent logarithmic behaviour, eg.
\be
O(L) = A_{LL}(aL^2) + L^{-1} A_{NLL}(aL^2) + L^{-2} A_{NNLL}(aL^2) + \cdots\;.
\ee
In this case, $r_n \sim L^{2n}$ but the preceding argument goes
through essentially unchanged, except that $\rho_n \sim L^{2n}$ as well.

Having obtained a resummed $\rho$ function, we can proceed to extract $\Lambda$ from the observable
making use of Eq.(\ref{eq:blogQ}) with $\mathcal{R}=O(L)/r_0(L)$.  In doing this, we may have
available {\it exact} values for $r_0$ and $r_1$ from a fixed-order calculation, which we can
use in place of their approximations from the resummed results.  This allows us to combine
resummed and fixed-order information in an essentially unique way (once we have
fixed the definition of our effective charge),  avoiding the so-called
``matching ambiguity'' associated with doing this in $\MSbPS$.
In particular the full exact NLO coefficient $r_1$ in a given RS is reproduced if
${\cal{R}}$ which solves \refeq{blogQ} is expanded in the coupling $a$ for that scheme, thanks to
the RS invariant ${\Lambda}_{\cal{R}}$ which appears on the lefthand side of the equation.

This provides us with a complete method for making ECH predictions including
resummations of large logarithms.  In the next section we test this approach by
comparing the distributions of thrust and heavy-jet mass in ${e}^{+}{e}^{-}$ annihilation
to NLL ECH predictions.

\section{ECH for Event Shapes at Next-to-leading Logarithmic Accuracy}
\label{se:ESV}

Event shape variables provide some of the most interesting and useful ways to confront QCD calculations
with experiment (for a recent review see \cite{Dasgupta:2003iq}). Their infrared and collinear safety
guarantees that they can be calculated in QCD perturbation theory, but fixed order calculations describe
their distributions rather poorly.  This situation can be improved by the recognition that for a shape variable $y$,
vanishing in the 2-jet limit, large logarithms $L=\log(1/y)$
appear at each order of perturbation theory and must be resummed.  However, a full understanding of the
observed distributions requires the introduction of large {\it non-perturbative} effects, {\it power corrections}
$\propto e^{-b/a} \simeq \Lambda/Q$ where $Q$ is some hard scale (e.g. the  $e^+ e^-$ centre of mass energy).
Although the existence of such effects can be motivated by considering simple models
of hadronization \cite{Webber:1994zd} or through a renormalon analysis \cite{Nason:1995hd}, their magnitudes are not at present calculable in a truly
systematic way from the QCD Lagrangian. Therefore to fit the data we require the introduction of either a phenomenological hadronization
model or additional non-perturbative parameters.  Although this hampers attempts
to extract reliable measurements of $\alpha_S$ from the data, it also provides a good opportunity to study the IR behaviour
of QCD experimentally.  For example, in Ref.\cite{Dokshitzer:1995zt} it was proposed to relate the magnitude of the
$1/Q$ power correction to event shape means to the average value of a hypothetical infrared-finite coupling.  This approach
leads to predictions for $1/Q$ power corrections to all event shape means in terms of a {\it single} additional parameter,
$\overline{\alpha}_0(\mu_I)$, the zeroth moment (i.e. mean) of the coupling at scales  $0<\mu<\mu_I$ (typically
$\mu_I \simeq 2$GeV).  In Ref.\cite{Dokshitzer:1997ew} it was shown how this approach could be extended to apply to event shape distrubutions.
Since then many experimental studies have appeared, fitting event shape means and distributions simultaneuosly for
 $\alpha_{\MSb}(M_Z)$ and $\alphazero(\mu_I)$.  Generally an approximate (up to corrections $\simeq 25\%$) universality of the $\alphazero$ values
is observed, supporting the hypothesis that power corrections can be related to a universal coupling in this way.  However, most
of these fits use the $\MSb $ scheme with the ``physical'' scale choice $\mu=Q$.
An exception is work within the Dressed Gluon Exponentiation (DGE) framework (see Refs.\cite{Gardi:1999dq,Gardi:2001ny,Gardi:2002bg} for the application to average thrust,
and the thrust and heavy-jet mass distributions).  For the event shape means another possible approach
is to work within the ECH framework.  This was first carried out for 1-thrust in
Ref.\cite{Campbell:1998qw}, and somewhat reduced power corrections were found compared to the physical scale approach.  This suggests the possibility that the
power suppressed effects are partly compensating for missing higher-order perturbative terms associated with the running of the coupling.
Indeed, similar conclusions were reached in the DGE approach, although there the subset of terms resummed differs significantly from that resummed
by the change of scale implicit in NLO ECH: the former is factorially divergent, the latter actually converges.   Recently
an analysis similar to that of Ref.\cite{Campbell:1998qw}, though more extensive,
was performed by the DELPHI collaboration \cite{r12}, taking into account effects arising from the finite bottom quark mass
via Monte Carlo simulations.  They found remarkably small power corrections within the ECH approach, which for many observables were
consistent with zero.  Indeed, the ECH predictions with {\it no} power corrections whatsoever gave a better description of the data
than the model of Refs.\cite{Dokshitzer:1995zt,Dokshitzer:1997ew} with a universal $\alphazero$.  In light of these surprising results, it is interesting to
consider applying the ECH method to the event shape distributions.

In fact, event shape distributions have previously been studied within the ECH framework \cite{r9,Abreu:2000ck}.
Ref.\cite{r9} in particular studied how the fit of the ECH results to data varied in quality for different regions of phase space.
To do this an effective charge was constructed separately for each bin of the data, and
NLO QCD calculations were used to extract $\Lambda_{\MSb}$ at centre of mass energy $Q=M_Z$.  The consistency of these
$\LMSb$ values between different data bins could then be examined.
Non-perturbative
effects were taken into account by using Monte Carlo simulations to correct the data back to ``parton-level''
distributions. This generally improved the consistency of the $\LMSb$ measurement, but with this approach it is hard to see
whether the ECH distributions prefer smaller hadronization corrections than the $\MSbPS$ ones.
Moreover, even after these corrections were applied there were still two kinematical
regions where the effective charge ceased to be a good description of the data, leading to instability in the $\LMSb$ values: the
2-jet limit where large logs enhance the higher-order perturbative coefficients, and the region (which exists
for many observables) where the LO result vanishes, causing $r_1 \rightarrow \infty$.  The latter problem
is hard to address within the effective charge approach, but the former problem seems to call for
a resummation of the effective charge beta-function, as described in Section \ref{se:resummation}.

In this section we study the distributions of 1-thrust ($\tau\equiv 1-T$) and heavy-jet mass ($\rho_h$).
We first show the effect of replacing the hadronization corrections of \cite{r9}
with an analytical power correction ansatz.
For simplicity, we use a shift in the distribution by an amount $C_1/Q$.
This form can be motivated by considering simple models of hadronization or through a renormalon analysis \cite{Webber:1994zd} and
has been found successful phenomenologically (see for example \cite{r12}).  Although better fits are often obtained using the model
of \cite{Dokshitzer:1995zt,Dokshitzer:1997ew}, because we are using a different perturbative approximation to standard NLO QCD, the
subtractions needed to remove double counting will not in general be so simple - in particular, it is not clear what scheme should
be used for performing the subtraction.

Next we consider placing the effective charge into the exponent of the integrated thrust distribution.  Even using
a NLO approximation for the effective charge, this has the
effect of resumming a series of logs in the distribution itself (in particular the ``double logs'' are included).

Finally we present results showing the effect of using the resummed $\rho$ functions described here in Section \ref{se:resummation}. 
First we investigate the extent to which higher order $\MSbPS$ logs are already included in the lower order
ECH predictions (``RG-predictability'').  Then we actually perform fits using the resummed ECH predictions.

For comparison, at all stages we also give results of fits to the same data using $\MSbPS$ (at NLO, LL and NLL accuracy).
As is customary, we use the variation of $\mu$ such that $Q/2 < \mu < 2Q$ to estimate a ``theoretical error''.

The general question of separating perturbative and non-perturbative effects also deserves comment.
Because perturbation theory diverges, it is not straightforward to define its sum; however, without doing this the magnitude of the ``non-perturbative''
effects is ambiguous.  Therefore, it is preferable to combine a fit for power corrections with a renormalon resummation, as in
\cite{Gardi:1999dq,Gardi:2001ny,Gardi:2002bg}.  We have not done so in this analysis, but as we are comparing ECH to $\MSbPS$
which differ only by a convergent set of higher order terms, we believe that our conclusions regarding the relative size
of power corrections stand.  It would of course be interesting to investigate the effects of a renormalon resummation on our ECH
results (ECH renormalon resummations have already been carried out for some single-scale observables in \cite{Maxwell:1996ig}).

Our data is taken over a wide range of centre-of-mass energies $Q=35-189$GeV (Refs.\cite{Abreu:2000ck},\cite{Buskulic:1992hq}-\cite{Braunschweig:1990yd}).  Lacking information on the correlation
between data points we have simply combined statistical and systematic errors in quadrature and performed a min-$\chi^2$ fit, allowing
$\chi^2$ to vary by 4 from its minimum to estimate a $2\sigma$ error.  This over-simplistic treatment means that our errors cannot
be considered reliable, however the central values of $\Lambda_{\overline{MS}}$ do give an impression of the effect of including the
power corrections and logarithmic resummation into the ECH framework.

Let us now consider the effects of analytical power corrections on the results of \cite{r9}.
The procedure used in \cite{r9} was to write an effective charge to represent the value of the event shape distribution
integrated over each bin of the data.  First, the Monte Carlo program EERAD \cite{Giele:1991vf} was used to compute the NLO perturbative coefficients
\footnote{ For our analysis we actually used EVENT2 \cite{Catani:1996jh} and we have checked that both programs give consistent coefficients.}
for each bin
\begin{equation}
\label{eq:NLOevtshape}
\int_{bin~i} dy ~ \frac{1}{\sigma} \frac{d\sigma}{dy}  = A_i \alpha + B_i \alpha^2 + O(\alpha^3).
\end{equation}
These were then used to write an effective charge, from which a value for $\Lambda_{\MSb}$ could be extracted by feeding the data into
\refeq{extracter}.  Here, to introduce a fit for $C_1$ we drop this ``direct extraction'' approach and instead
perform a minimum $\chi^2$ fit for $\Lambda_{\MSb}$ and $C_1$.  For this to work, we need to exclude the regions where the EC approach
cannot fit the data.  For comparison with \cite{r9}, we choose the same ranges selected there (based on
the flatness of $r_1$), except that the lower end of the range is made proportional to $1/Q$ when looking at data away from $Q=M_Z$.
The reason for this is that sub-leading non-perturbative effects are expected to become important for $y \simeq \Lambda/Q$ \cite{Dokshitzer:1997ew,Korchemsky:1995zm,
Korchemsky:1999kt}.
As we are shifting the predictions before comparing to data we require the NLO coefficients evaluated for arbitrary bin edges.  We have
approximated these using a set of order 6 polynomial interpolations from the output of EVENT2.  We have checked by halving the Monte Carlo bin size to 0.005
that this induces no sizeable error (using the doubled sampling changes the best fit values here by less than 2\%).

\begin{figure}
\begin{center}
\includegraphics[scale=0.5]{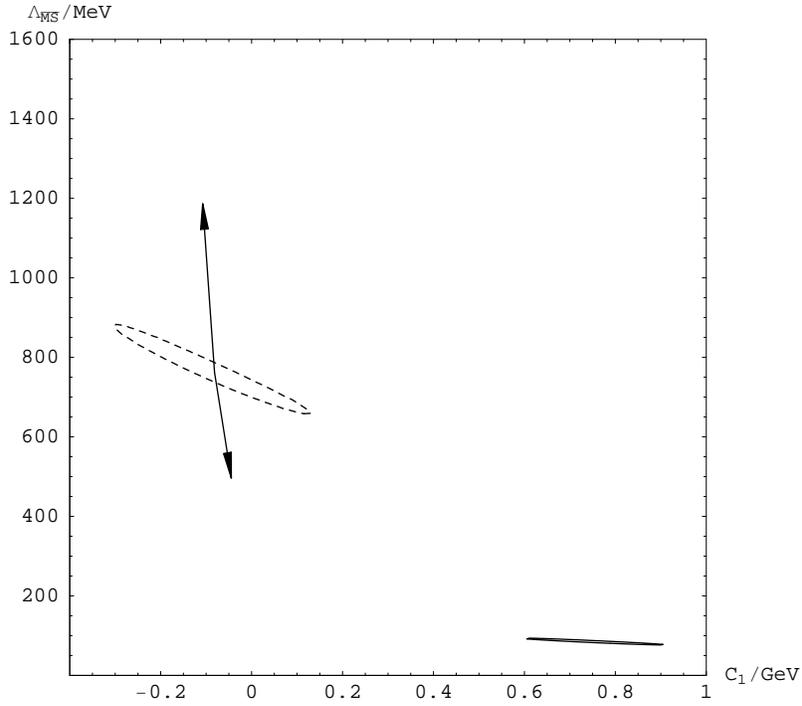}
\end{center}
\caption{1-thrust: Fits for $\Lambda_{\MSb}$ and $C_1$
within the framework of \cite{r9} (solid ellipse) and standard NLO QCD peturbation theory (dashed ellipse).
In the latter case the scale is chosen to be $\mu=Q$, and the effect on the central value of a change of renormalization scale by a factor of 2 is indicated by the arrows.
2$\sigma$ error contours are shown (from allowing $\chi^2$ to vary within 4 of its minimum).  The fit range is $1-T=0.055M_Z/Q-0.23$.}
\end{figure}

\begin{figure}
\begin{center}
\includegraphics[scale=0.5]{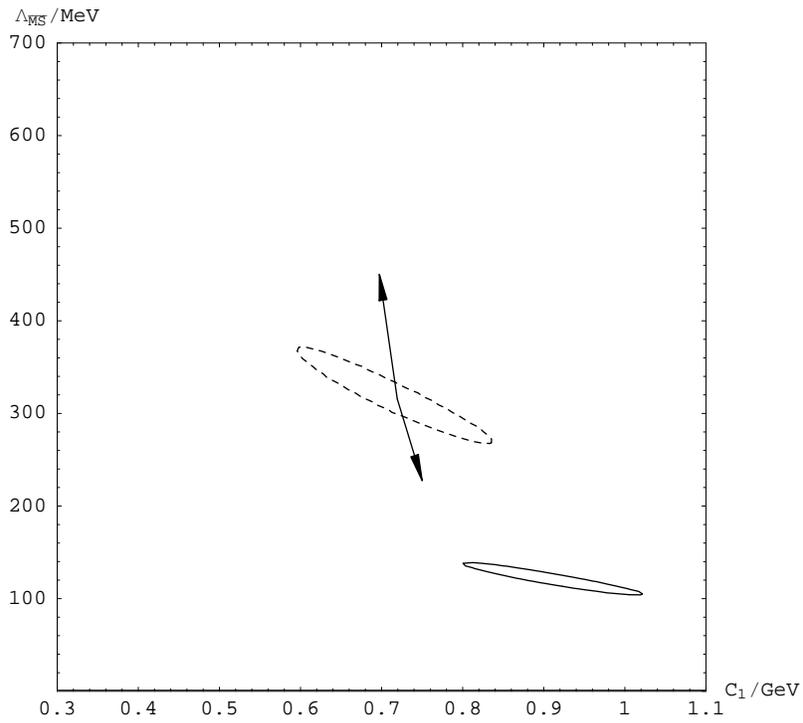}
\end{center}
\caption{As Fig. 1 but for heavy-jet mass.  The fit range is $\rho_h=0.035M_Z/Q-0.2$. }
\end{figure}

The results for 1-thrust and heavy-jet mass are presented in Fig. 1 and
Fig. 2 respectively.

In the case of thrust it appears that the ECH results prefer larger power corrections (and significantly smaller $\LMSb$ values).
For heavy-jet mass, the situation is similar, although the differences are not quite as extreme.  However, in
both cases we find comparable $\Lambda_{\MSb}$ values to those found in \cite{r9} using hadronization corrected data rather than
an analytical power correction ansatz.
\footnote{
  However, the values of $\Lambda_{\MSb}$ quoted in \cite{r9} are actually wrongly normalized for two reasons.  Firstly the factor of
  $(2c/b)^{(c/b)} \simeq 0.85$ was omitted, so the results are really values for $\tilde{\Lambda}_{\MSb}$.  Secondly, the results
  of EERAD were normalized to the {\it Born} cross-section $\sigma_0$, whereas the data are normalized to the {\it total} cross-section
  $\sigma$, and this was not taken into account.  Mutiplying the EERAD perturbation series by a correction
  factor $\sigma_0/\sigma=1-\alpha/\pi+\cdots$ decreases $r_1$ by exactly 1,  increasing the extracted $\Lambda_{\MSb}$ values
  by $e^{1/b}$.  So the total correction factor to apply to the results of \cite{r9} is $(2c/b)^{(c/b)}e^{1/b} \simeq 1.11$.
}

A crucial property that an event shape must possess in order for a resummation of logarithms to be feasible
with present techniques is so-called exponentiation.  To illustrate this property, consider
the typical form of an event shape distribution as a double expansion in $a$ and $L=\log(1/y)$:
\begin{equation}
\label{eq:non-exponentiated}
\frac{1}{\sigma} \frac{d\sigma}{dy} = A_{LL}(aL^2) + L^{-1} A_{NLL}(aL^2) + L^{-2} A_{NNLL}(aL^2) + \cdots\;.
\end{equation}
The $A$ functions have a perturbative expansion $A(x) = {A}_{0}x+{A}_{1}{x}^{2}+\ldots$ and for $\tau$ and $\rho_h$ are
known up to NNLL accuracy.  If the event shape exponentiates, then we can also write
\begin{eqnarray}
\label{eq:exponentiated}
R_y(y') \equiv \int_0^{y'} dy \frac{1}{\sigma} \frac{d\sigma}{dy} = C(a\pi) \exp(L g_1(a\pi L) + g_2(a\pi L) + a g_3(a\pi L) + \nonumber \\
a g_3(a\pi L) + \cdots) + D(a\pi ,y)\;.
\end{eqnarray}
For $\tau$ and $\rho_h$,  $g_1$ and $g_2$ are known \cite{Catani:1991kz}.  When working with this form of the distribution it is conventional to refer to
$g_1$ as containing the {\it leading logarithms} and $g_2$ as containing the {\it next-to-leading logarithms}.
$C = 1+O(a)$ is independent of $y$, and $D$ contains terms that vanish as $y \to 0$.  These can be calculated to NLO by comparison
with fixed-order results.  However, there is no unique way of including this fixed-order information into $R_y(y')$ (this is the
so-called matching ambiguity).  For example, it is also legitimate to include the $C, D$ terms into the exponent (termed ``log R matching''),
as the difference is of order $a^3$.

In \refeq{non-exponentiated}, there are terms at ${\rm O}(a^n)$ with up to $2n$ factors of $L$ multiplying them.  
In contrast, in the exponent of \refeq{exponentiated} at ${\rm O}(a^n)$ we find no more than $n+1$ factors of $L$.
The $L^ma^n$ terms with $n+1 < m \leq 2n$ are generated by the exponentiation.  For example, including just the leading order,
leading log term $\sim L^2a$ in the exponent of \refeq{exponentiated} leads, after the exponent is expanded out, to the entire
set of double-logs $\sim L^{2n}a^n$ in \refeq{non-exponentiated}.  Ideally we would like to use this exponentiation
property in our ECH approximation.
So, let us consider the effect of defining
\begin{equation}
\label{eq:expECH}
R_y(y') = \exp(r_0(y') \mathcal{R}(y'))\;.
\end{equation}
Here all the physics is encoded into a single effective charge, which is exponentiated
in its entirety.
This is similar to log R matching in that if we re-expand $r_0 \cal{R}$ in terms
of $a$ and $L$ the $C$ and $D$ functions will clearly appear in the exponent.
However, in this approach there is no matching ambiguity because
once we have picked the effective charge the inclusion of $C$ and $D$ is automatically determined. 
The function $r_0$ for thrust can be found by integrating
the analytically known leading order 1-thrust distribution \cite{PinkBook}
\be
\left. \frac{1}{\sigma} \frac{d\sigma}{d\tau} \right |_{LO}=\frac{C_F a\,\left( 3 - 9\,\tau  - 3\,{\tau }^2 + 9\,{\tau }^3 + \left( -4 + 6\,\tau  - 6\,{\tau }^2 \right) \,\log ( \frac{1}{\tau } - 2) \right)} {2\, \,\tau \left(  \tau-1  \right) }
\ee
with the boundary condition that $R_{\tau}$ vanishes to LO for $\tau \geq 1/3$.  The result is
\ba
r_0(\tau)&=& C_F \left( -\frac{5}{4}  + \frac{{\pi }^2}{6} + 3\,\tau  + \frac{9\,{\tau }^2}{4} + \left( \frac{3}{2} - 3\,
\tau  \right) \,(\log (1 - 2\,\tau )-\log(\tau)) \right. \nonumber \\
&& \left. -   {\left( \log (1 - \tau ) - \log (\tau ) \right) }^2 -   2\,{\rm Li}_2 \left ( \frac{\tau }{1 - \tau } \right ) \right ).
\ea
1-Thrust and heavy-jet mass agree at LO, so the same result holds for heavy-jet mass.
A given prediction for $\R(y)$ now allows us to calculate a corresponding $R(y)$.  Then,
by taking the difference in $R(y)$ across the
bins in each experimental data set, a comparison to data can be carried out, including a $1/Q$ power correction by using
$R_{PC}(y) = R_{PT}(y - C_1/Q)$.
In the remainder of this paper we will consider predictions for the distributions of 1-thrust and heavy-jet mass
arising from subsituting various ECH approximations into \refeq{expECH}.

The simplest possibility is to use a standard NLO ECH $\R(y)$.  This only requires knowledge of $r_1$, which can be easily
obtained from the results of Monte Carlo calculations of the distributions to NLO.
This NLO ECH, re-expanded in $a(Q)$ and $L$ in the $\MSb$ scheme, includes terms $\sim L^m a^n$ for all $n$ and all $m \leq n+1$.
Some of these terms can be compared with their exactly known LL and NLL counterparts allowing us to determine to
what extent the LL and NLL terms are ``RG-predictable''.
In this context, RG-predictability refers to the extent to which the higher order coefficients
in the perturbation series for $\R$ are already present in some lower order ECH result.  Overall, the logs are not very well
predicted by the NLO ECH results except for rather small n (see Figs. 3 and 4 for NLL examples).  
However, the exponentiation of the effective charge
will produce further towers of logs in the distributions themselves.
We might therefore expect these NLO ECH results to have better behaviour
in the 2-jet region than the results of Ref.\cite{r9}. This is indeed the case, as can be seen in Fig. 5.
It is interesting to note that exponentiation of a NLO $\MSbPS$ series in place of $\R$ produces a 
distribution (the dashed curve in Fig. 5) with a badly misplaced peak (this remains the case for any reasonable value of $\LMSb$).
Therefore, the good qualitative description of the peak is only obtained at NLO with the use of both
exponentiation and ECH (until we introduce non-perturbative effects).
\footnote
{
  At NLO, ECH is equivalent to a scale choice $\mu = Q e^{-r_1/b}$.  In the case of both thrust and
  heavy-jet mass $r_1 = bL/2 + {\rm const} + \cdots$, so using a NLO ECH
  is equivalent to choosing $\mu = Q \sqrt{y} f(y)$ where $f(y)$ goes to a constant as $y \to 0$.
  This is interesting as a ``physical scale'' argument where one takes, for example, the heavy-jet mass $m_h=\sqrt{\rho_h}Q$
  as the scale would give $\mu = Q \sqrt{\rho_h}$.  So these two scale-setting methods have the same leading $\rho_h$ dependence,
  and only differ by the factor $f$.
  This factor is important, however, as its overall normalization is RS-dependent and this ensures that $\mu$ is
  chosen in such a way that we obtain the same ECH answer whatever RS we choose.
  
}

\begin{figure}
\begin{center}
\includegraphics[scale=0.6]{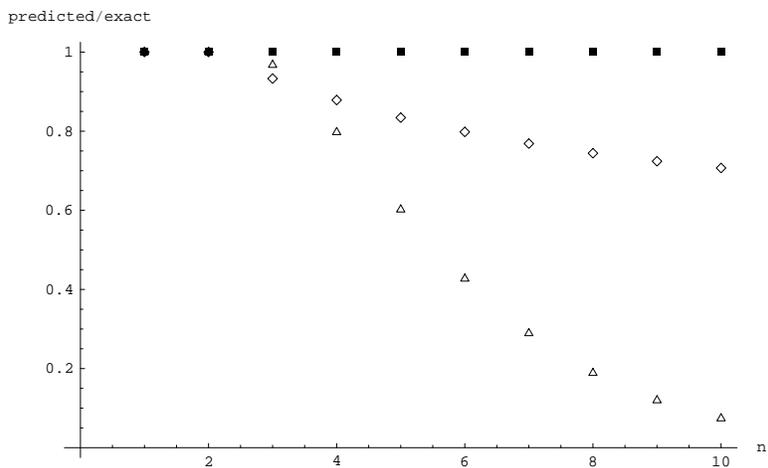}
\end{center}
\caption{Prediction of NL $\MSbPS$ logs in the exponent for 1-thrust based on re-expanding lower order ECH results in $a_{\MSb}$.
 The ratio of the predicted NL cofficient at ${\rm O}(a^n)$
with the corresponding exact coefficient is shown.  Triangles are for NLO ECH, diamonds for LL ECH. As a consistency check, squares show the results
for NLL ECH where the NL log must appear exactly.}
\end{figure}

\begin{figure}
\begin{center}
\includegraphics[scale=0.6]{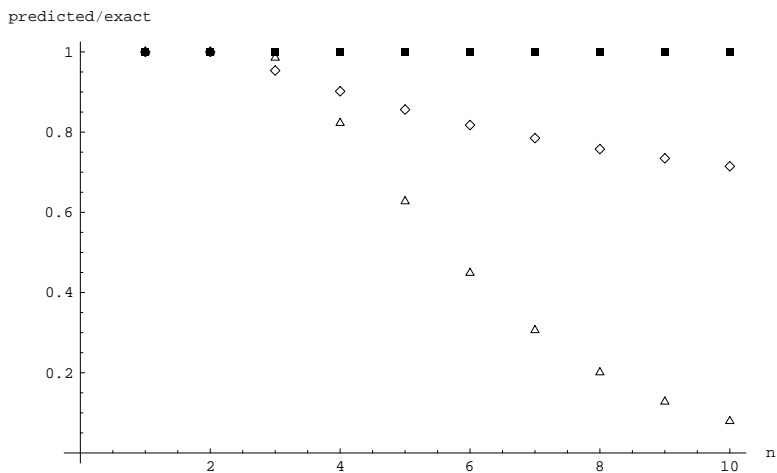}
\end{center}
\caption{As Fig. 3 but for heavy-jet mass.}
\end{figure}

\begin{figure}
\begin{center}
\includegraphics[scale=0.7]{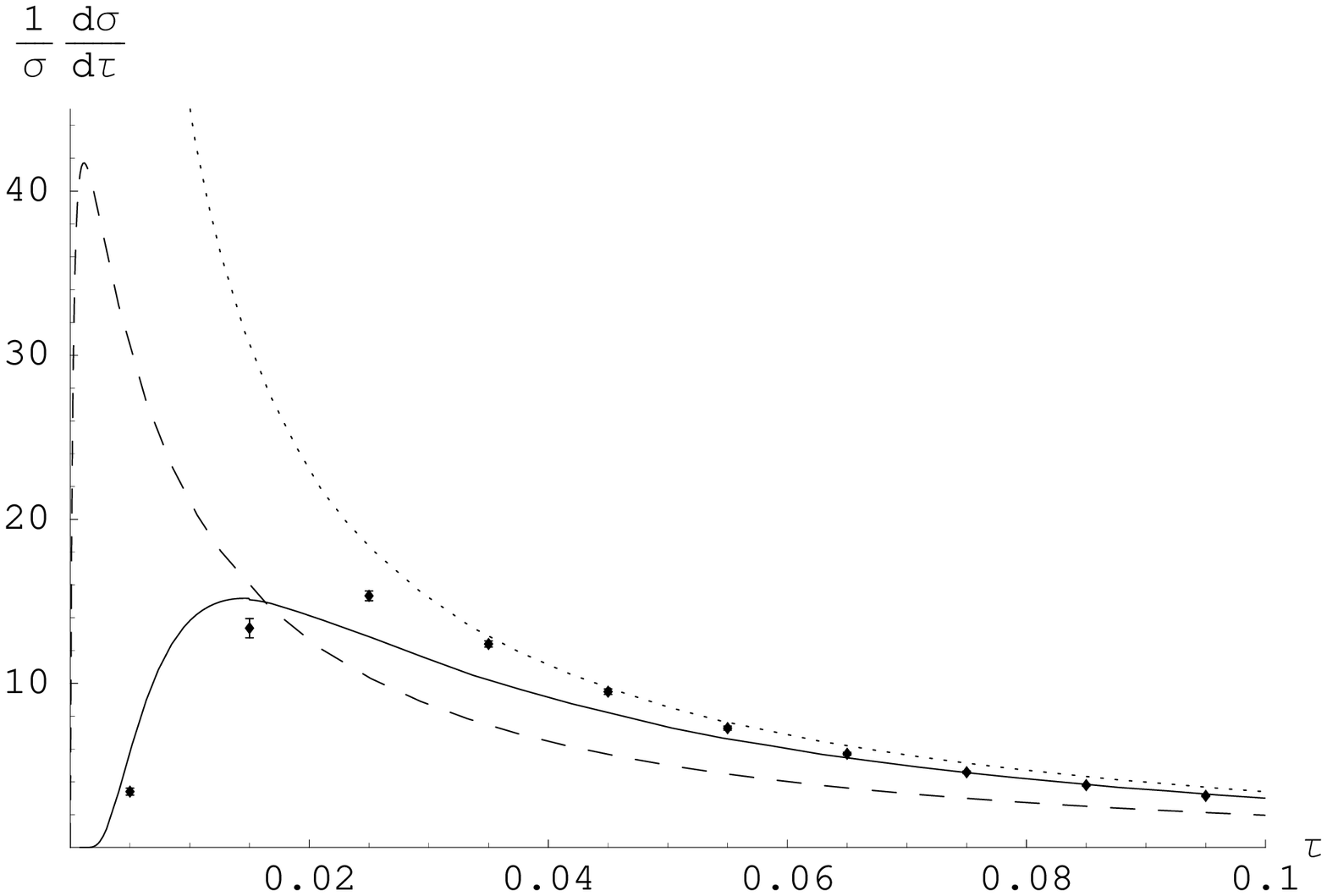}
\end{center}
\caption{Comparison of the 1-thrust distributions calculated using various NLO approximations in the 2-jet region.  The solid curve arises from the exponentiated ECH of
\refeq{expECH}.  The dashed curve is obtained by expanding this effective charge in the $\MSb$ scheme with $\mu=Q$.  The dotted curve
is a prediction in the approach of Ref.\cite{r9}.  Throughout we have taken $Q=M_Z$ and $\LtMSb=250$MeV.  For comparison, DELPHI data at $Q=M_Z$ are shown.}
\end{figure}

We can now consider performing a resummation of logs in the $\rho$ function
as described in Section \ref{se:resummation}.  
First, constructing these order-by-order in $\R$
allows us to again address the question of the ``RG-predictability'' of the
$\MSbPS$ logs.  
For example, one can ask how much of the NL log at ${\rm O}(a^n)$ in $\MSbPS$ is included
when we use the LL ECH result (of course, if we use the NLL ECH the full NL $\MSbPS$ log is included by construction).
To find out, we can re-expand the ECH results in terms of $\alpha_{\MSb}(Q)$ as we did for the NLO ECH.  The resulting coefficients are shown
in Fig. 3 (for 1-thrust) and Fig. 4 (for heavy-jet mass) as fractions of the exact coefficients.  The LO and NLO
($n=1,2$) coefficients agree exactly as $r_0$ and $r_1$ have been used to NLL accuracy in all the predictions.  There
is a clear improvement in the prediction of the NL logs as we move from NLO ECH to LL ECH as one might expect.  The
extent to which the NL logs really are included in LL ECH is encouraging, as it suggests that NLL ECH might do
a good job of including some higher order corrections that are omitted in the $\MSbPS$ approach.

The method described in Section \ref{se:resummation} can now be used to produce numerical approximations to 
$\rho_{\rm LL}$ and $\rho_{\rm NLL}$ (these are not truncated at any order in $\R$).
All of our calculations were carried out using
the computer algebra system {\it Mathematica} \cite{Mathematica}, allowing the use of arbitrary precision arithmetic
in taking the $L \to \infty$ limit.
These $\rho_{\rm LL}$ and $\rho_{\rm NLL}$ functions can be used to make predictions for $\R$, by inserting them
in \refeq{blogQ} and numerically solving
the transcendental equation.  To ensure exact cancellation between the singularities in
Eq.(\ref{eq:curlyG}), for $\mathcal{R}<0.005$ we use the exact series expansions of $\rho(\mathcal{R})$
up to to order $\mathcal{R}^4$ (the difference is totally negligible for all values of $L$ we consider)
This defines what we call our ``LL ECH'' and ``NLL ECH'' predictions.

To perform fits with these exponentiated effective charges we again need to select a fit range.
After exponentiation, the problem as $r_0 \to 0$ remains, and in fact
for thrust worsens; unfortunately this means we need to restrict the fits to $1-T < 0.18$, $\rho_h < 0.24$  to obtain good fits
in the ECH approach. Irrespective of the inclusion of logs, the onset of non-perturbative effects
more complicated than a simple $1/Q$ shift means that we still need to impose a lower cut.  These higher-order
non-perturbative effects are expected to be of order $\Lambda/(Qy)$ so our cut should be placed at $y \sim \Lambda/Q$
with $\Lambda$ some infrared scale.
One might expect the inclusion of the extra logs into $\rho$ to improve the fit of the ECH prediction to data in the 2-jet region.
Unfortunately, it turns out that including these logs actually {\it worsens} the behaviour
of the ECH results.  In this region the growth of $r_1$ causes $\R$ to become larger
(because $\Lambda_{\mathcal{R}}$ approaches $Q$), and this is accelerated by the addition of logs into $\rho$.
In fact, $\mathcal{R}$
eventually grows large enough that we encounter a branch cut in $\rho$ which appears due to the branch cut in $g_1$ \cite{Catani:1991kz}.
Clearly this behaviour is unphysical, and must be avoided in our fits to data.
Presumably some other higher order corrections
intervene to produce an ECH prediction which is well-behaved in the 2-jet limit.
In any case, in light of the RG-predictability of the sub-leading $\MSbPS$ logs, it is still possible
that the $\rho$ resummations improve the quality of the ECH predictions in the intermediate region,
so it is worth trying to fit the data with a lower cut in place (and studying the sensitivity of the best
fit parameters to the choice of this cut).
Good fits are obtained over the whole energy range using $\rho_h,1-T>0.05M_Z/Q$.

Any data bins not lying within the range $0.05M_Z/Q<\rho_h<0.24$, $0.05M_Z/Q<1-T<0.18$ have been left
out of the fit; a summary of the data we actually used is given in Tables 3 and 4. We have also removed the JADE data at $35$ and
$44$ GeV from the heavy-jet mass fits, since its inclusion dramatically worsens the fit
quality for both the $\MSbPS$ and ECH predictions.
The result of fitting for $\LMSb$ and $C_1$ are shown in Figs. 6 and 7 for all three approximations.  For comparison,
$\MSbPS$ results are also shown.
The fit range for these could in principle be extended as they do not suffer
from the $r_0 \to 0$ problem that afflicts the ECH, however, in order to facilitate a direct comparison between the two approaches
we have used the same fit range for both.  The most notable
feature of the results is the stability of the $\Lambda_{\overline{MS}}$ values found within the ECH
framework as we move from NLO to LL and then to NLL accuracy, while the fit quality hardly changes.
It shows that the effect of the leading and next-to-leading logarithms can be mimicked by an increase in $C_1$ (which itself
is not large).  This improved stability with respect to the order of the approximation might be a consequence of the
RG-predictability of the $\MSbPS$ logs discussed above, because, for example, a lot of the logs that only turn up at NLL order in
the $\MSbPS$ predictions are included already at LL order in ECH.
It must be noted however that despite their stability, these $\Lambda_{\overline{MS}}$ values are still smaller than the world average.  
Some examples of the actual NLL ECH distributions are shown in Fig. 8.

To investigate the sensitivity of these results to our choice of fit range we have redone the fits for a ``low'' range and
a ``high'' range.  The low range was determined by decreasing the upper cut until half the bins were excluded, and the
high range was determined by increasing the lower cut similarly.  The effects of these changes on the central values
of $\Lambda_{\overline{MS}}$ and $C_1$ are shown in Table 1 (for 1-thrust) and Table 2 (for heavy-jet mass).
The ECH fit values for heavy-jet mass appear more stable than the $\MSbPS$ ones; and the stability increases
as the accuracy of the predictions are increased.  This is also true for the thrust, but to a lesser extent.  A particular exception is that the 
application of NLL ECH to the ``high'' fit range gives a significantly different power correction compared to the ``low''
fit range.  This is probably responsible for the relatively large $\chi^2$ for the ``normal'' fit.
The reason for the change in $C_1$ may be the $r_0 \to 0$ problem being exacerbated by the increase in size of the effective charge as more
logs are added into its beta-function.  Because the $r_0 \to 0$ problem represents a breakdown of our approximations, the ``low'' fit range
results are probably more trustworthy (and in any case, agree very well with the ``normal'' fit range results).

Lastly, we have also considered the so-called ``modification of the logs'' that is often invoked in studies of event shape
variables.  This consists of modifying $L=\log(1/y) \to \log((2y_{max}-y)/y)$ to ensure that the resummed parts of the expression
vanish at the upper kinematic limit $y_{max}$ (which is 0.5 for both $T$ and $\rho_h$).   The change in central values is shown in Tables 1 and 2. One finds that the fitted values change
very little. This is to be expected since the restricted fit range automatically
ensures that the logarithm is essentially unchanged in that region.

\begin{table}
\begin{tabular}{|l|c|c|c|}
\hline
Prediction & $\Lambda$/MeV & $C_1$/GeV & $\chi^2/dof.$ \\
\hline
NLO ECH & 98,116,118 & 1.02,0.85,0.89 & 24/46,59/88,27/44\\
NLO $\MSbPS$ & 446,463,517 & 1.47,1.42,1.28 & 26/46,57/88,23/44\\
\hline
LL ECH & 101,119,108 & 0.80,0.63,0.88 & 23/46,63/88,30/44\\
LL $\MSbPS$ & 371,417,478 & 1.12,1.02,0.83 & 24/46,49/88,22/44\\
\hline
NLL ECH & 103,107,93 & 0.50,0.52,0.93 & 23/46,84/88,34/44 \\
NLL ECH (mod) & 104,110,95 & 0.50,0.47,0.90 & 23/46, 81/88, 34/44 \\
NLL $\MSbPS$ & 200,233,268 & 0.94,0.79,0.60 & 23/46, 49/88, 23/44\\
\hline
\end{tabular}\\
Table 1.  Sensitivity of our fit values for $\Lambda_{\overline{MS}}$ and $C_1$ to the choice
of fit range for thrust.  The fit ranges are $0.05M_Z/Q-0.1$ (low), $0.05M_Z/Q-0.18$ (normal), $0.11M_Z/Q-0.18$ (high).
\end{table}

\begin{table}
\begin{tabular}{|l|c|c|c|}
\hline
Prediction & $\Lambda$/MeV & $C_1$/GeV & $\chi^2/dof.$\\
\hline
NLO ECH & 120,114,142 & 1.19,1.22,0.94 & 19/42,50/84,29/41 \\
NLO $\MSbPS$ & 236,115,221 & 1.65,1.84,1.27 & 19/42,68/84,37/41\\
\hline
LL ECH &  124,123,128 & 1.06,1.07,1.01 & 21/42,51/84,29/41\\
LL $\MSbPS$ & 185,132,146 & 1.39,1.57,1.33 & 19/42,63/84,36/41\\
\hline
NLL ECH & 125,127,122 & 0.99,0.97,1.04 & 21/42,51/84,29/41\\
NLL ECH (mod) & 126,128,122 & 0.98,0.97,1.05 & 21/42, 51/84, 29/41 \\
NLL $\MSbPS$ & 114,82,67 & 1.29,1.49,1.62 & 19/42,64/84,37/41 \\
\hline
\end{tabular}\\
Table 2.  Sensitivity of our fit values for $\Lambda_{\overline{MS}}$ and $C_1$ to the choice
of fit range for heavy-jet mass.  The fit ranges are $0.05M_Z/Q-0.12$ (low), $0.05M_Z/Q-0.24$ (normal), $0.125M_Z/Q-0.24$ (high).
\end{table}

\begin{figure}
\begin{center}
\includegraphics[scale=0.6]{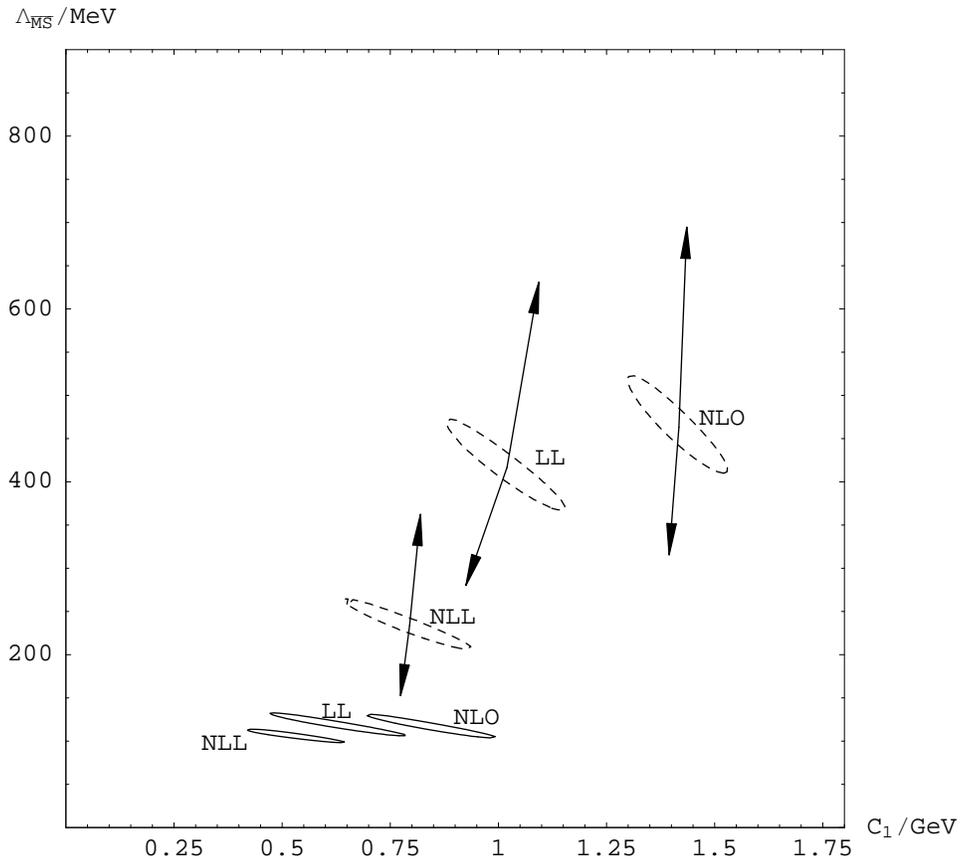}
\end{center}
\caption{Fits to the thrust distribution for $\Lambda_{\MSb}$ and $C_1$.  Solid ellipses use ECH, dashed ellipses $\MSbPS$
(with the arrows showing the effect on the central value of varying $Q/2<\mu<2Q$). The ellipses indicate 2$\sigma$ errors generated by allowing
$\chi^2$ to vary within 4 of its minimum.  For a summary of the data used see Table 3. }
\end{figure}

\begin{figure}
\begin{center}
\includegraphics[scale=0.6]{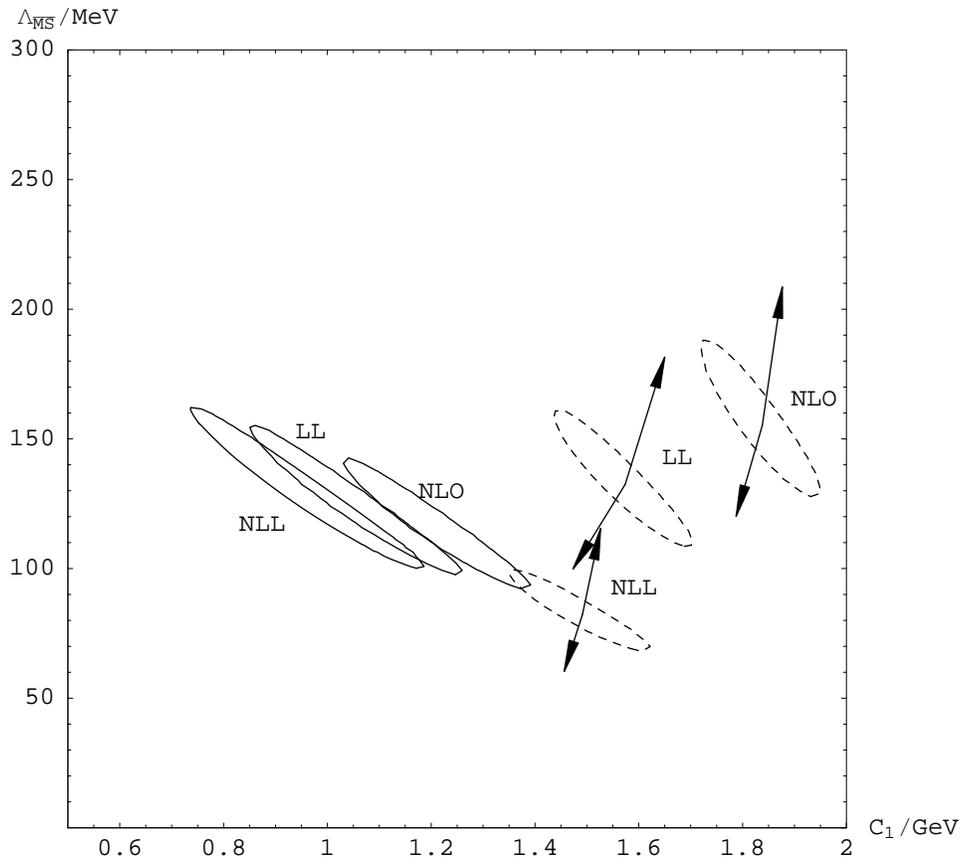}
\end{center}
\caption{As Fig. 6 but for heavy-jet mass.  For a summary of the data see Table 4. }
\end{figure}

\begin{figure}
\begin{center}
\includegraphics[scale=0.5]{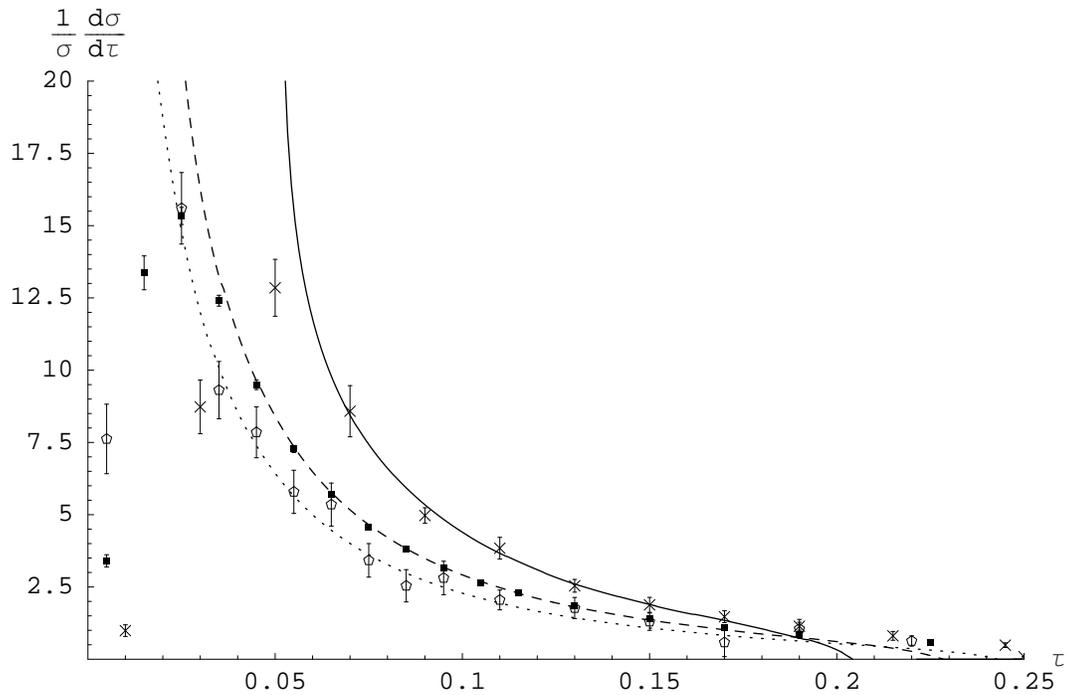}
\end{center}
\caption{Examples of our best fit NLL ECH 1-thrust distributions. The solid curve is for $Q=44$GeV, dashed is $Q=91.2$GeV and dotted is $Q=183$GeV.
These predictions are compared to data from the JADE and DELPHI collaborations at these energies;  crosses are JADE data at 44GeV, boxes are DELPHI data
at 91.2GeV and open circles are DELPHI data at 183GeV.}
\end{figure}

\begin{table}
\begin{tabular}{|l|c|c|c|c|}
\hline
Experiment & Q & Range & Data Points & Source\\
\hline
ALEPH & 91.2 & 0.05-0.18 & 7 & \cite{Buskulic:1992hq}\\
 & 133 & 0.04-0.15 & 4 & \cite{Buskulic:1996tt}\\
\hline
DELPHI & 91.2 & 0.05-0.18 & 10& \cite{Abreu:2000ck} \\
 & 133 & 0.04-0.18 & 5 & \cite{Abreu:1999rc}\\
 & 161 & 0.04-0.18 & 5 & \cite{Abreu:1999rc}\\
 & 172 & 0.04-0.18 & 5 & \cite{Abreu:1999rc}\\
 & 183 & 0.03-0.18 & 11 & \cite{Abreu:1999rc}\\
\hline
JADE & 35 & 0.14-0.18 & 2 & \cite{MovillaFernandez:1997fr}\\
 & 44 & 0.12-0.18 & 3 & \cite{MovillaFernandez:1997fr}\\
\hline
L3 & 91.2 & 0.065-0.175 & 4 & \cite{Adeva:1992gv}\\
 & 189 & 0.025-0.175 & 6 & \cite{Acciarri:2000hm}\\
\hline
OPAL & 161 & 0.03-0.15 & 6 & \cite{Ackerstaff:1997kk}\\
 & 172 & 0.03-0.15 & 6 & \cite{Abbiendi:1999sx}\\
 & 183 & 0.03-0.15 & 6 & \cite{Abbiendi:1999sx}\\
 & 189 & 0.03-0.15 & 6 & \cite{Abbiendi:1999sx}\\
\hline
SLD & 91.2 & 0.06-0.16 & 3 & \cite{Abe:1994mf}\\
\hline
TASSO & 44 & 0.12-0.16 & 1 & \cite{Braunschweig:1990yd}\\
\hline
\end{tabular}\\
Table 3. Summary of the data used in our fits for thrust.
\end{table}

\begin{table}
\begin{tabular}{|l|c|c|c|c|}
\hline
Experiment & Q & Range & Data Points & Source \\
\hline
DELPHI & 91.2 & 0.05-0.2 & 8 & \cite{Abreu:2000ck}\\
 & 161 & 0.04-0.2 & 4 & \cite{Abreu:1999rc}\\
 & 133 & 0.04-0.2 & 4 & \cite{Abreu:1999rc}\\
 & 172 & 0.04-0.2 & 4 & \cite{Abreu:1999rc}\\
 & 183 & 0.03-0.24 & 10 & \cite{Abreu:1999rc}\\
\hline
SLD & 91.2 & 0.08-0.24 & 3 & \cite{Abe:1994mf}\\
\hline
ALEPH & 91.2 & 0.05-0.2 & 8 & \cite{Buskulic:1992hq}\\
\hline
L3 & 91.2 & 0.051-0.216 & 7 & \cite{Adeva:1992gv}\\
 & 189 & 0.03-0.24 & 14 & \cite{Acciarri:2000hm} \\
\hline
OPAL  & 91.2 & 0.0625-0.2025 & 4 & \cite{Acton:1992fa}\\
 & 161 & 0.0289-0.2025 & 5 & \cite{Ackerstaff:1997kk} \\
 & 172 & 0.0289-0.2025 & 5 & \cite{Abbiendi:1999sx} \\
 & 183 & 0.0289-0.2025 & 5 & \cite{Abbiendi:1999sx} \\
 & 189 & 0.0289-0.2025 & 5 & \cite{Abbiendi:1999sx}\\
\hline
\end{tabular}\\
Table 4.  Summary of the data used in our fits for heavy-jet mass.
\end{table}

An alternative to the approach followed here would be to apply our resummations to an effective charge associated to the
distribution itself similar to the one considered in Ref.\cite{r9} but taking the bin width to 0 (the NLO approximation for this
effective charge is shown as the dotted curve on Fig. 4).
This is certainly possible and leads to a $\rho$ function where the LL, NLL and NNLL terms are known.
However this creates problems with the resummation which prevent this approach from yieding useful predictions.
To see this, consider $\rho_{\rm LL}$.  As explained in Section \ref{se:resummation}, this
is related to the leading log $\bar{\rho}$ via
\be
\rho_{\rm LL}(\mathcal{R})=\bar{\rho}_{\rm LL}(\mathcal{R}) + bc \mathcal{R}^3.
\ee
In this (unexponentiated) case $\bar{\rho}_{\rm LL}$ is found using the one-loop beta-function and the double logarithmic distribution
\be
\frac{1}{\sigma} \frac{d\sigma}{dy} = \frac{d}{dy} \exp (- k L^2 a ) = (2 k e^{-L} L) a \exp (-k L^2 a) \equiv ( 2 k e^{-L} L) \bar{\mathcal{R}}\;,
\ee
where $k$ is a constant ($4/3$ for thrust and heavy-jet mass).  This distribution has a peak as a function of $a$ at $a_{max}=1/kL^2$, and so its
inverse only exists for $\bar{\mathcal{R}} < \bar{\mathcal{R}}(a_{max}) = e^{-1}/kL^2$.  As a consequence $\bar{\rho}_{LL}(\mathcal{R})$ vanishes
at this point (where a branch cut starts).  Adding the $bcR^3$ term to give $\rho_{\rm LL}$, and later adding the NLL terms, does not remove this branch cut.
As $\mathcal{R}$ is evolved from $Q=\infty$ it increases until it reaches this maximum value, and then its evolution becomes undefined.
This value turns out to be too small to allow fits to the data.  One could possibly ``switch branches'' of $\rho$ at this point
and allow $\mathcal{R}$ to decrease again, although this would
of course still not provide a good fit to the data.  Note also that this zero of $\bar{\rho}$ does {\it not } correspond to an ``infrared freezing'' type
behaviour because $\bar{\rho}$ approaches the zero as $(\mathcal{R}-\mathcal{R}_{max})^\gamma$ with $\gamma<1$ - thus the singularity in \refeq{blogQ} is integrable
and the zero is reached after a finite amount of evolution in $Q$.

\section{Discussion and Conclusions}
\label{se:conclusions}

In Section \ref{se:resummation} we showed how it was possible to build a resummation of large infra-red logarithms
into ECH.  In principle, this allows the use of ECH for multi-scale observables, even
when the ratio between the scales grows large.  The approach taken was to apply a resummation
at the level of the effective charge beta-function $\rho$.  A method was described allowing
such a resummed $\rho$ to be extracted numerically to any desired accuracy from the resummed
expression for an observable in $\MSbPS$.

In Section \ref{se:ESV} we used the results of Section \ref{se:resummation} to extend the direct extraction of ${\Lambda}_{\overline{MS}}$ from
${e}^{+}{e}^{-}$ event shape observables of Ref.\cite{r9} to include a resummation of
large infra-red logarithms (in this case $L={\log}(1/y)$ where $y$ is the shape variable). One could relate the observable
${R}_{y}({y}^{\prime})$ to an effective charge ${\cal{R}}$ by
exponentiation, ${R}_{y}({y}^{\prime})={\exp}({r}_{0}{\cal{R}})$. Using the NLO approximation for $\R$
lead to a good fit to data.
One could then numerically
construct ${\rho}_{LL}({\cal{R}})$ and ${\rho}_{NLL}({\cal{R}})$ functions
by resumming to all-orders the corresponding pieces of the ${\rho}_{n}$ in \refeq{blogQ}. The LL and NLL
predictions for ${R}_{y}({y}^{\prime})$ for a given value of ${\Lambda}_{\overline{MS}}$ then follow on inserting these
${\rho}({\cal{R}})$ functions in \refeq{blogQ} and numerically solving the transcendental equation.
To model $1/Q$ power corrections we fitted to a shifted distribution ${R}_{PC}(y)=R_{PT}(y-{C}_{1}/Q)$.
Whilst in principle straightforward a number of complications arose. In particular as $1-T$ approaches
$1/3$ the leading coefficient $r_0$ goes to zero, invalidating the effective charge approach. This places
a rather stringent upper limit on the fit range.
There are also problems in the two-jet region due to the growing $\R$ running into a branch cut in $\rho$
which appears as the image of a branch cut in $g_1$ in the $\MSbPS$ approach. We also noted that one
cannot directly relate the observables to an unexponentiated
effective charge, as in Ref.\cite{r9}, since in that case ${\rho}_{LL}({\cal{R}})$ has a different branch cut such that the energy evolution of
${\cal{R}}$ becomes undefined and we are unable to fit the data.
Simultaneous fits for ${\Lambda}_{\overline{MS}}$ and ${C}_{1}$ were performed using data
for thrust and heavy-jet mass distributions over a wide range of energies (see Tables 3 and 4).
The $2{\sigma}$
error contours in ${\Lambda}_{\overline{MS}}$ and ${C}_{1}$ are shown in Figs.6 and 7. NLO,
LL and NLL results are shown for both standard $\MSbPS$, and
for ECH. For $\MSbPS$ there is a strong decrease in ${\Lambda}_{\overline{MS}}$
going from NLO to LL to NLL, whereas for ECH the fitted ${\Lambda}_{\overline{MS}}$ values
are remarkably stable. The fitted value of ${C}_{1}$ is also somewhat smaller for ECH.
We also investigated the stability of the fits to changing the fit range in Tables 1 and 2. The
ECH results show more stability than $\MSbPS$.
This, along with the stability of the fitted values with respect to the order of approximation,
leads us to believe that the fit ranges we have chosen are restrictive enough to avoid
the problems that appear in the 2-jet region for LL and NLL ECH, whilst hopefully retaining the benefit of
including some RG-predictable sub-leading logs into our predictions.
The fits all produce $\Lambda_{\MSb}$ values somewhat smaller than the world average.
It is interesting to note, however, that
they are of the same magnitude as those found in \cite{r9}.  Converting our NLL best fit $\Lambda$ values to
$\alpha_{\MSb}(M_Z)$ using the 2-loop beta-function
gives $\alpha_{\MSb}(M_Z)=0.106$ for $\Lambda_{\MSb}=100$MeV (thrust)  and $\alpha_{\MSb}(M_Z)=0.109$ for $\Lambda_{\MSb}=125$MeV
(heavy-jet mass).  Similarly small 
values of $\alpha_{\MSb}$ have also been reported in the DGE approach (see \cite{Gardi:2002bg}).  It is possible
that sizeable NNLO corrections (omitted by both the DGE and ECH resummations) might be responsible; it will be
interesting to see the effect of matching to fixed-order NNLO results when they become available.  
\\

We would conclude that, notwithstanding the limited fit range and technical complications from
which ECH suffers, there is evidence that the fitted power corrections are reduced
relative to the standard approach, although not as dramatically as in the DELPHI fits of
Ref.\cite{r12} which are consistent with zero power corrections. However in that analysis
corrections for bottom quark mass effects were made, which were not included in our analysis.
Event shape means have also been measured in DIS at HERA \cite{Herameans}, and it would be interesting
to perform an ECH analysis in that case as well. Unfortunately, 
in this case it is not known how to construct an ECH approximation for DIS where one has
a convolution of parton distributions and hard scattering QCD cross sections. One can, however,
apply PMS \cite{Stevenson:1981vj} to choose the renormalisation and factorisation scales. An analysis along these lines
is in progress \cite{mjd}.

\section*{Acknowledgements}

M.J.D. gratefully acknowledges receipt of a PPARC UK Studentship.

\end{document}